\documentclass[mathleft
]{an}
\usepackage{graphicx}
\usepackage{times}
\begin{document}

\Pagespan{789}{}
\Yearpublication{2006}%
\Yearsubmission{2005}%
\Month{11}%
\Volume{999}%
\Issue{88}%

\title{Status of the Cherenkov Telescope Array project}

\author{Ulisses Barres de Almeida\inst{1}\fnmsep\thanks{On behalf of the CTA Consortium}
\fnmsep\thanks{Corresponding author:
  \email{ulisses@cbpf.br}\newline}
}
\titlerunning{Cherenkov Telescope Array}
\authorrunning{U. Barres de Almeida}
\institute{
Centro Brasileiro de Pesquisas Físicas, Rua Dr. Xavier Sigaud 150, 
22290-180 Rio de Janeiro, Brasil
}

\received{30 May 2005}
\accepted{11 Nov 2005}
\publonline{later}


\abstract{%
Gamma-ray astronomy holds a great potential for Astrophysics, Particle Physics and Cosmology. The CTA is an international initiative to build the next generation of ground-based gamma-ray observatories, which will represent a factor of 5-10$\times$ improvement in the sensitivity of observations in the range 100 GeV - 10 TeV, as well as an extension of the observational capabilities down to energies below 100 GeV and beyond 100 TeV. The array will consist of two telescope networks (one in the Northern Hemisphere and another in the South) so to achieve a full-sky coverage, and will be composed by a hybrid system of 4 different telescope types. It will operate as an observatory, granting open access to the community through calls for submission of proposals competing for observation time. The CTA will give us access to the non-thermal and high-energy universe at an unprecedented level, and will be one of the main instruments for high-energy astrophysics and astroparticle physics of the next 30 years. CTA has now entered its prototyping phase with the first, stand-alone instruments being built. Brazil is an active member of the CTA consortium, and the project is represented in Latin America also by Argentina, Mexico and Chile. In the next few months the consortium will define the site for installation of CTA South, which might come to be hosted in the Chilean Andes, with important impact for the high-energy community in Latin America. In this talk we will present the basic concepts of the CTA and the detailed project of the observatory. Emphasis will be put on its scientific potential and on the Latin-American involvement in the preparation and construction of the observatory, whose first seed, the ASTRI mini-array, is currently being constructed in Sicily, in a cooperation between Italy, Brazil and South Africa. ASTRI should be installed on the final CTA site in 2016, whereas the full CTA array is expected to be operational by the end of the decade.}

\maketitle

\section{The TeV sky}
After decades of technical struggle, it was in 1989 that the first very-high energy (VHE, $E>100$ GeV) gamma-ray source, the Crab Nebula, was finally observed at TeV energies, opening up what would become the last astronomical window into the cosmos, and the prime observational technique for the study of the extreme universe. The breakthrough came with the Whipple Collaboration (Weekes et al. 1989), which operated a pioneer instrument on Mt. Hopkins, Arizona. The discovery was only possible thanks to the image analysis techniques developed by Michael Hillas, from the University of Leeds, which allowed to finally single out the tiny gamma-ray signal from the swarm of cosmic-ray background events that hit the atmosphere every second (Hillas, 1985), at a rate approximately 300 Hz.

The field then progressed steadily but slowly, with the detection of Markarian 421 as the first extragalactic VHE source coming up in 1992 (Punch et al. 1992). Until the turn of the Century and the entering in operation of the current generation of ground-based instruments, namely H.E.S.S.\footnote{https://www.mpi-hd.mpg.de/hfm/HESS/}, MAGIC\footnote{https://magic.mpp.mpg.de} and VERITAS\footnote{http://veritas.sao.arizona.edu}, not much more than a dozen of VHE sources where known. 

Today, after a decade of operation of these facilities, ground-based gamma-ray astronomy has undergone a revolution, and is now solidly established as a new area of astronomy, as well as the most productive experimental branch of astroparticle physics. The VHE sky today numbers over 150 sources, over 50 of which are extragalactic, and in average every month a new object is added to the list.\footnote{See tevcat.uchicago.edu} In the Galaxy, the sources emitting VHE photons are of almost all kinds, with dozens of supernova remnants, pulsar-wind nebulae, molecular clouds, as well as a few binary star systems, stellar clusters and even pulsars being detected. Dark sources, with still unknown counterparts in other wavelengths are also numerous and testimony to the potential of the field in being at the forefront of astronomical discovery (de Naurois, 2013).

In the extragalactic sky, blazars figure as the most abundant source class, aided by the relativistic Doppler-boosted emission from its jets. In fact, the science of Active Galactic Nuclei (AGNs) and the study of relativistic outflows has been completely renewed by gamma-ray observations at the VHE window (Lorenz \& Wagner, 2012). 

An extrapolation of the success of the field between the year 2004 and today into the future suggests that within the the next decade the TeV sky, thanks to the work of the planned CTA, will be shining with over a thousand sources, distribute around the entire Galaxy and over cosmological distances, as we will detail below (Funk, 2012).

\section{Detecting VHE gamma-rays}
The science of VHE gamma-ray astronomy is intimately connected to the physics of cosmic-rays, since these highly energetic particles are the ones responsible for the very-high energy emission seen from the astrophysical sources. VHE gamma-ray astronomy could therefore be described as the "imaging of cosmic particle accelerators". The direct detection of the cosmic ray particles would undoubtedly be the ideal way to probe the acceleration sites, through the simultaneous knowledge of their energy and composition. Nevertheless, the Galactic magnetic field acts on the charged particles of the cosmic rays, deviating their trajectories and effectively isotropising their reconstructed arrival directions, erasing any directional information about their point of origin in the sky. Only for the most energetic of the comic ray particles, with $E \sim 10^{20}$ eV, could we hope to obtain some information about their true origin in the sky, but the fluxes are too small for a cosmic ray astronomy to effectively develop (see e.g., Pierre Auger Collab. 2013). Gamma-ray observations enter therefore the scene of astroparticle physics as an important ally, allowing for the direct imaging of the sources of comic rays in the Galaxy and beyond. 

There are essentially two general ways in which gamma-rays can be produced by cosmic rays in astrophysical sources. The first, leptonic process, is by means of inverse Compton scattering, by which VHE electrons up scatter low energy photons over a broad energy range above the initial one, by a factor $\propto \gamma^2$, the Lorentz factor of the scattering electrons. In the presence of magnetic fields, energetic electrons will unavoidably produce lower-energy radiation via the synchrotron process, and therefore it is natural that in many sources the up scattered photons will come from this very population, in what is called Synchrotron Self-Compton mechanisms. Aditionally, in environments with a high density of low energy ambient photons, the so-called external-Compton can also contribute to the emission.

VHE gamma-ray photons can also be produced by hadronic processes, when accelerated protons or nucleons interact with ambient matter to produce pions

\begin{equation}
  p + {\rm nucleus} \rightarrow p' ... + \pi^{+/-} + \pi^{0} + ...~, 
\end{equation}  
 
\noindent which then decay through the following channels, producing gammas and neutrinos:
 
 \begin{equation}
  \pi^{0} \rightarrow 2\gamma ~; ~ \pi \rightarrow \mu\nu_{\mu} ~;~  \mu \rightarrow e \nu_{mu} \nu_{e}
\end{equation}

From these considerations it is apparent that whereas gamma-ray observations allow for the direct imaging of the cosmic accelerators, the direct detection of the sites of cosmic ray production is a more complex business since they require that observations be able to distinguish between hadronic and leptonic processes and this distinction remains challenging to date, depending essentially on an improvement of the sensitivity of the observations, to which CTA will be abel to respond adequately. Of course, the direct observations of neutrinos could nail down the hadronic nature of the accelerator, but their detection is even more challenging at the moment (Aartsen et al., 2014).

\subsection{Current IACTs and the status of the field}
The pioneer gamma-ray instruments\footnote{Namely, Whipple, in the US, and CAT and HEGRA in Europe; see Weekes, T., 2008, for a brief review by one of the pioneers of this early age of gamma-ray astronomy} which operated during the 90s and were responsible for the development of the imaging analysis techniques, stereoscopy and fast-electronic cameras which were necessary to launch the field of VHE Astronomy were simply not sensitive enough (limit sensitivity of $\sim$ 0.1 Crab) to firmly detect much more than a handful of TeV sources. These were nevertheless sufficient to hint at the potential contributions of the field to relativistic astrophysics, and early in 2000 there began the plans for construction of the current, third, generation of instruments, designed to reach sensitivity of 1\% Crab after 50 hours on integration time.

One of the principal milestones achieved by the new generation of instruments was the successful completion by H.E.S.S. of the Galactic plane scan. Making use of their large, 5$^{\circ}$  FoV, H.E.S.S. surveyed the entire galactic plane from $l = 260^{\circ} ~{\rm to}~ 65^{\circ}$ in galactic longitude and $|b| < 3.5^{\circ}$ in latitude, for 2800 hours of observations over 10 years, discovering more than 60 sources down to approximately 10 mCrab in sensitivity (see Carrigan, S. et al., 2013 and references therein). The survey revealed that sources of VHE gamma-rays are ubiquitous in the Galaxy and among the many different classes of astrophysical objects. For large extended sources such as some nearby supernova remnants, it has been possible to investigate some of the objects sub-structures (e.g., Aharonian, F. et al., 2006).

As for the extragalactic sky, the sources are dominated by Active Galactic Nuclei (AGN), specially blazars, whose detection is aided by the Doppler boosted emission from the relativistic jets. In fact, the VHE gamma-ray data has reignited research on AGNs, discovering extremely energetic emission and fast variability processes from a number of these objects, much beyond what was expected beforehand (e.g., Aleksic et al. 2014). Likewise the study of nearby radio galaxies with ground-based instruments, together with radio telescopes and multi-band satellites turned out to be one of the best relativistic astrophysics laboratories available to investigation (VERITAS Collab. et al., 2009). The emission from distant BL Lacs (limiting distance today is shortly below $z \sim 1$) has also been extensively used to study the extragalactic background light (EBL; Abramowski et al. 2013b) to search for axions (Abramowski et al. 2013a) and violation of the Lorentz invariance principle (Barres de Almeida \& Daniel, 2012), as well as a means to achieve the difficult task of measuring the intensity of the intergalactic magnetic field IGMF (Arlen, T. et al., 2012)

Finally, more recently, with the accumulation of data and the improved sensitivity of the current instruments after a series of upgrades (Aleksic et al., 2015; Bomont, J. et al. 2014), it has been possible to indirectly probe, with relatively good sensitivity, for dark matter annihilation signatures using VHE gamma-ray observations. Results do not hint at any distinguishable signal from DM, but stringent constraints start to be derived for particles with masses up to a few TeV (Abramowski, A. et al., 2015). Despite the continuos efforts, and MAGIC's capability of fast-repositioning of its 70-ton telescope structure to any point in the sky in 20 seconds, as well as the low-energy threshold of 30-40 GeV of the fifth, 27-m class H.E.S.S. telescope (Bolmont, J. et al. 2014), no gamma-ray bursts have yet been detected at VHE energies and it seems the long-awaited discovery will be relegated to CTA (Aleksic et al. 2014). 

\section{The Cherenkov Telescope Array}
The Cherenkov Telescope Array (CTA) Consortium is a major project to build the next generation ground-based gamma-ray observatory (Actis et al. 2011), aiming to dramatically boost the current imaging atmospheric Cherenkov telescope (IACTs) performances and widen the VHE science case (Acharya et al., 2013). The principal characteristic of CTA, which distinguishes it from previous experiments in the field is that it will be the first ever ground-based gamma-ray open observatory, meaning that CTA will work essentially on the basis of observation proposal submission and data access will be open to the entire astronomical community. This new mode of operation will represent a major change in the way VHE Gamma-ray Astronomy operates and will be the key factor that will boost science in the field.

Another unique, key feature of CTA is that it will be composed of two sites, one in the Northern Hemisphere and one in the South, so to provide full sky coverage. On each site a few dozen IACTs will deployed, which will work in stereoscopic mode to provide high-sensitivity observations over a wide energy range, aimed to go from as low as 20 GeV to hundreds of TeV. 

The large number of instruments, circa 100 telescopes in the South and about 30 in the Northern site, distributed respectively over 10 km$^2$ and 3 km$^2$, will provide a large effective area to the observatory, required to allow it to achieve the desired sensitivity even at the highest energies, where photon fluxes are very low. The requirement of a wide spectral coverage, spanning over 5 orders of magnitude in energy, imply that the observatory must be composed of three different classes of instruments, dubbed Large, Medium and Small Size Telescopes -- respectively, LSTs, with 23 m in diameter;  MSTs, 12 m in diameter; and SSTs 4-5 m in diameter (Vercellone et al. 2014). 

Each telescope class will follow a specific design, optimised for a different energy band, he LSTs being responsible for the lower energy extension below 100 GeV (Ambrosi et al., 2013) and the more numerous and least expensive SSTs, scattered over the largest ground area, responsible for detection of the highest energy photons up to hundreds of TeV (Fasola et al. 2014). The different instrument "sub-arrays" will be integrated as most as possible, to facilitate working in conjunction, by the application of common components and solutions whenever possible, such as with the cameras, mirrors and control systems, for example (e.g., Oya et al. 2014). Additionally, a new class of Schwarzschild-Couder dual-mirror telescopes (SCTs), with primary mirror size of $\sim 9.5$ m, will also be added to the Southern array to further improve the angular resolution of the observatory in the key energy range between  0.1-10 TeV (Cameron 2012). 

Whereas the Northern site will most likely be composed of a few LSTs and 20-30 MSTs, the Southern site will count with additional 50-70 SSTs (plus 20-30 SCTs) to respond to particular requirements associated to the Galactic science case, e.g., to extend the energy range of observations above 100 TeV for the study of PeVatrons, as well as to  achieve a good angular resolution required for imaging the structure of extended sources (Acero et al. 2013).

Apart from the energy range extension, CTA will provide an order-of-magnitude improvement in overall flux sensitivity, approaching the 1 mCrab level (for 50 hours integration), and achieving an angular resolution as low as 0.1$^\circ$-0.05$^\circ$ between 0.1-1 TeV, which will have an important impact on the surveys of the Galactic plane (Dubus et al. 2013). The large field of view of the telescopes, of $\sim5^{\circ}$ will open way for a possible use of  CTA for survey of a few specially elected regions of the extragalactic sky and serendipitous discoveries. An exceptional energy resolution of 10\% at 1 TeV, and better than 25\% throughout the energy range of observations will allow for unprecedented spectral reconstruction, with impact in a number of items of the science case.

\begin{figure*}
\includegraphics[width=\textwidth]{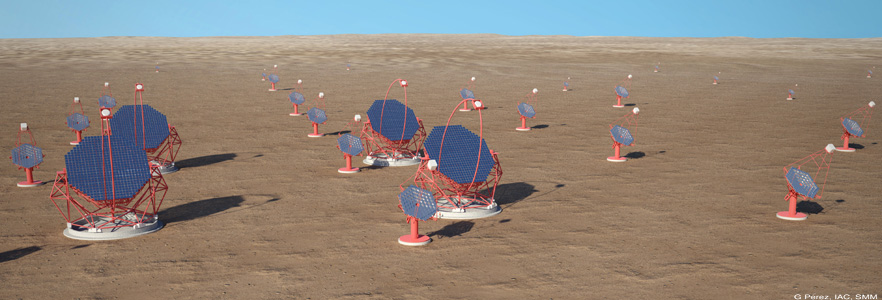}
\caption{Artist conception of the future Cherenkov Telescope Array, showing a probable layout for the different telescope classes.}
\label{cta}
\end{figure*}

\subsection{From current arrays to CTA}

When gamma-rays of energies above $\sim 5$ GeV enter the atmosphere, they decay in the vicinity of the atoms of the air to produce pairs, in a process which initiates a leptonic cascade that propagates down, inducing the emission of Cherenkov light. The ground signal this radiation produces, which is observed by the IACTs, has the form of a circular pool of Cherenkov light (for normal incidence) of about 250 m in diameter, seen as a short pulse of a few ns duration. Therefore, in order to work stereoscopically, a few IACTs  (ideally 3 or 4, to avoid loss of sensitivity in a specific direction) must be placed in arrangements within an area in the light pool, so that coincidence triggers of a same shower can be frequent. This is the case of current ground-based arrays.

When we move from this to CTA, the effective area of the array is not anymore given by the $10^{5} m^{2}$ of the Cherenkov light pool of single showers, but by the ground area covered by the telescopes, which for CTA will be of $\sim$ 10 km$^2$. At this point, a new issue arises which refers to the optimisation of the array layout, both in terms of the number of telescopes to be deployed and their relative arrangement. In the case of CTA the situation is more complex, because the optimisation must be made over a very large energy range, combining different kinds of telescopes, and Monte Carlo simulations of different array configurations are currently being conducted to find the  design for optimising CTA science (Bernloehr et al. 2013).

\subsection{CTA requirements and performance}
The success of CTA will be associated to it achieving the key design goals defined for it, based on the science case prepared for the observatory (CTA Consortium 2011), namely: (i) a 10-fold increase in sensitivity at TeV energies; (ii) a 10-fold increased effective energy coverage; (iii) larger field of view, $\sim$ 8$^\circ$, to perform surveys (Dubus et al. 2013; Funk \& Hinton 2013); (iv) substantially improved angular resolution; (v) and full sky coverage, achieve by having two observatories, one at each hemisphere.

As shown in figure 2, with such a configuration, the sensitivity at high energies will be limited solely by the ground area covered by the array, since the photon fluxes above 10 TeV are quite low. At lower energies, the observations will be background limited, with the aggravating fact that at very low energies, of tens of GeV, systematic effects in the identification of showers and the process of gamma/hadron separation start to be quite strong and limit the potential of the observations. 

Apart from these general requirements, there are some specific aspects of the design which CTA should meet to boost its discovery potential. Performance at the extreme limits of the observational range, both at low and at high energies, will be quite important. At low energies, the strong absorbing effects of EBL mean that a decade improvement in energy, say from 30 GeV to 20 GeV might make quite a big difference for the detection of high-z sources (Sol et al. 2013; Mazin et al. 2013). Likewise, at high energies, the successful study of PeVatrons depends on pushing the observational limits up to 300 TeV (Acero et al. 2013). 

Detailed morphological studies of the nearest extended sources will require the achievement of a few arcmin angular resolution (de Ona-Wilhelmi et al. 2013), whereas a 10\% resolution in energy is estimated to be necessary should we wish to detect spectral features and lines such as one expects to see for Dark Matter decay/annihilation signatures (D'oro et al. 2013). Finally, to achieve another of its main science goals, which is the detection of VHE emission of gamma-ray bursts (GRBs), CTA will need the capability to rapidly slew at any position in the sky in the shortest time possible, and the LSTs are being designed to perform such a task within a 20-second window (Inoue et al. 2013).

The successful performance of CTA will depend, as discussed in the previous section, not only on the individual instruments' design, but also on the configuration of the ground array as a whole.

\begin{figure}
\includegraphics[width=82mm]{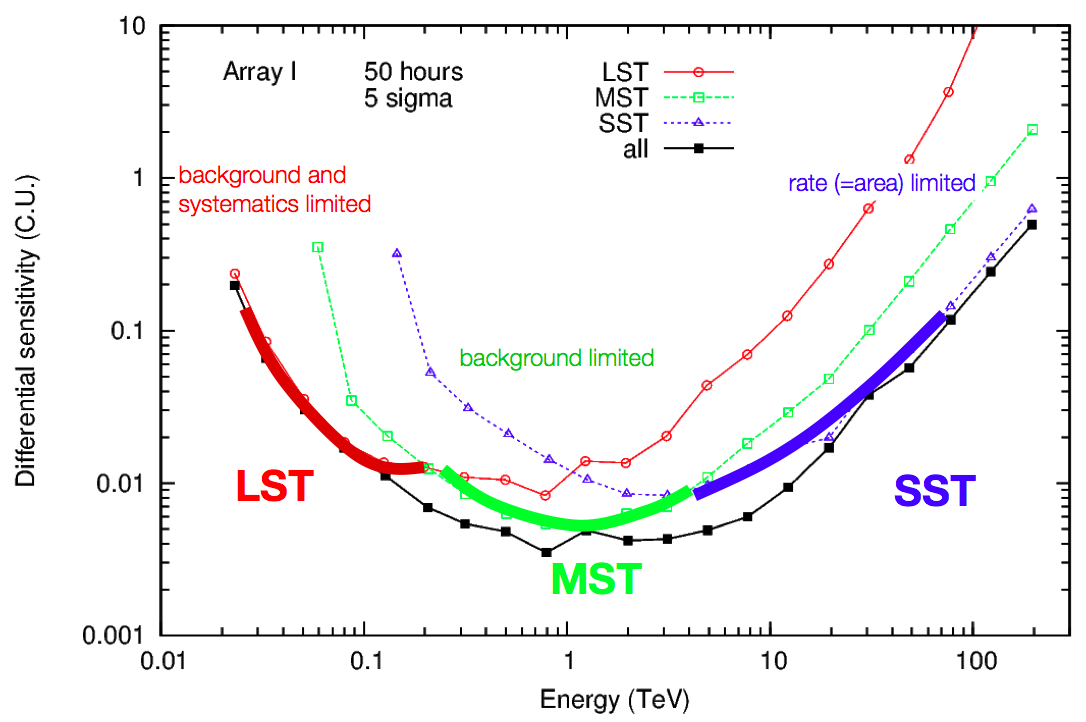}
\caption{Prospective CTA sensitivity curve for the configuration "Array I" (Bernloehr et al. 2013), showing how contribution of different telescopes combine to achieve the observatory goals.}
\label{cta_curves}
\end{figure}


\subsection{The CTA telescopes}
As previously mentioned, due to its wide energy range and broad science case, CTA will be composed of four different telescope types. Despite this complexity, a fundamental aspect of the project, which in a sense secures its success, is that most of the designs are based on proven technology developed for and applied in the current generation of ground-based gamma-ray isntruments. The one-mirror telescopes of large (Ambrosi et al. 2013), medium (Behera et al. 2012) and small (Moderski et al. 2013) sizes are in fact close variants of the traditional Davies-Cotton designs used in H.E.S.S., MAGIC and VERITAS, and historically adopted in the field due to its optical isochronicity properties, adequate for the short, nanosecond-duration pulses of the shower's Cherenkov emission (in detriment, nevertheless of some other optical properties such as a small PSF). CTA is currently in prototyping phase, and the first models of each of the three instruments are being built for preliminary tests, so that once the location of the observatory is defined, installation on site can commence.

Despite widely used and successful, the Davies-Cotton design is incompatible with a low cost solution to attain all of CTA's main goals. For this reason an innovative, dual-mirror, Schwarschil-Couder (SC) design has been developed, which is able o achieve small pixel size, large FoV and improved angular resolution, all simultaneously and at a low cost (Vassiliev et al. 2007). In this configuration, which will be adopted in an alternative version of the small size telescopes, the focal plane is located between two aspherical mirrors, close to the secondary mirror. Apart from this, other innovations that which are being studied for CTA regard the light detectors, with silicon-based photo-multipliers (SiPMs) being considered for some of the designs, which would allow smaller pixel cameras as well as moonlight operation, thus increasing the observatory duty-cycle (Biland et al. 2014). 

The ASTRI project (Pareschi et al. 2014, Fiorini et al. 2014), an Italian initiative to build the first SC-design IACT, is the first CTA prototype to have had its first light, earlier this year.  The full, end-to-end prototype has been built in 2014 and installed in the slopes of the Etna Volcano in Sicily, and is currently under test. Once the CTA site is defined, and as soon as basic site infrastructure is ready, the 7 telescopes which will compose the ASTRI mini-array will be installed on site as a precursor array to the larger CTA.


\subsection{Where to build CTA}
CTA is currently well-inside its prototyping phase, with all four telescope classes having their first test instruments under construction. During this year and in 2016, the operation of the designs will be evaluated and it is expected that in 2017 on-site installation will start. In the next months, the site for both the Northern and the Southern arrays will therefore be defined, so that the CTA project will finally enter construction phase. Currently in each Hemisphere a shortlist of two preferential sites has been made: in the South, a choice will be made between an ESO site in Cerro Armazones, in the high altitudes of the Chilean Andes and the high plateaus of Namibia, at Aars. For the North, definition will be between the MAGIC telescope site at La Palma, in the Canary Islands and a site in Northern Mexico.



\acknowledgements
We acknowledge the support of FAPERJ through a grant {\it Projeto Temático} 110.148/2013 for funding the participation of CBPF in the CTA project.    



\begin{thebibliography}{}
  \bibitem{} Aartsen, M.G. et al. (IceCubeCollab.): 2014, PRL 113, 101101 	
  \bibitem{} Acero, F. et al.: 2013, APh 43, 276.
  \bibitem{} Acharya, B.S. et al. (CTA Consortium): 2013, APh 43, 3
  \bibitem{} Actis, M. et al. (CTA Consortium): 2011, Exp. Astron. 32, 193
  \bibitem{} Abramowski, A. et al. (H.E.S.S. Collab.): 2013a, Phys Rev. D 88
  \bibitem{} abramowski A. et al. (H.E.S.S. Collab.): 2013b, A\&A 550, 4 
  \bibitem{} Abramowski, A. et al. (H.E.S.S. Collab.): 2015, PRL 114, 081301
  \bibitem{} Aharonian, F. et al. (H.E.S.S. Collab.): 2006, ApJ 636, 746.
  \bibitem{} Aleksic, J. et al. (MAGIC Collab.): 2013, MNRAS 437, 3103
  \bibitem{} Aleksic et al. (MAGIC Collab.): 2014, Science 346, 1080
  \bibitem{} Aleksic, J. et al. (MAGIC Collab.): 2015, APh in press.
  \bibitem{} Ambrosi, G. et al.: 2013, Proc. 33rd ICRC, arXiv:1307.4656
  \bibitem{} Arlen, T. et al. (VERITAS Collab.): 2012, ApJ 757, 123
  \bibitem{} Barres de Almeida, U. \& Daniel, M.K.: 2012, APh 35, 850
  \bibitem{} Behera, B. et al.: 2013, Proc. of SPIE 8444, 17
  \bibitem{} Bernloehr, K. et al.: 2013, APh, 43, 171
  \bibitem{} Biland, A. et al.: 2014, NIMA 766, 15
  \bibitem{} Bolmont, J. et al.: 2014, NIM A 761, 46
  \bibitem{} Cameron, R.A.: 2012, SLAC-PUB-15122
  \bibitem{} Carrigan, S. et al.: 2013, Proc. 33rd ICRC, arXiv:1307.4690 
  \bibitem{} CTA Consortium: 2011, Exp. Astron. 32, 193
  \bibitem{} de Naurois, M.: 2013, AdSpR 51, 258
  \bibitem{} de Oã-Wilhelmi, E. et al.: 2013, APh. 43, 287 
  \bibitem{} Dubus, G. et al.: 2013, APh 43, 317
  \bibitem{} Fasola, G. et al.: 2014, Proc. SPIE 9145, 914551
  \bibitem{} Fiorini, M. et al. 2014, Proc. SPIE 9150, 915024
  \bibitem{} Funk, S.: 2012, Proc. 31st ICRC arXiv:1204.4529 
  \bibitem{} Funk, S. \& Hinton, J.A.: 2013, APh. 43, 348
  \bibitem{} Hillas, A.M.: 1985, Proc. 18th ICRC 3, 445
  \bibitem{} Inoue, S. et al.: 2013, APh. 43, 252
  \bibitem{} Lorenz, E. \& Wagner, R.: 2012, EPJ H 37, 459
  \bibitem{} Mazin, D. et al.:2013, APh 43, 241
  \bibitem{} Moderski, R. et al.: 2013, Proc. 33rd ICRC, arXiv:1307.3137
  \bibitem{} Oya, I. et al..: 2013, Proc. SPIE 9152, 91522G
  \bibitem{} Pareschi, G. et al.: 2014, Proc. HEAD Meeting 14, 116.25
  \bibitem{} Pierre Auger Collab.: 2013, ApJL 762, L13
  \bibitem{} Punch, M. et al. (Whippler Collab.): 1992, Nature 358, 477
  \bibitem{} Sol, H. et al.: 2013, APh. 43, 215.
  \bibitem{} VERITAS Collab. et al.:2009, Science 24, 444
  \bibitem{} Vassiliev, V.V. et al.: 2013, APh 28, 10
  \bibitem{} Vercellone, S. et al. (CTA Consortium): 2014, NIMA 766, 73.
  \bibitem{} Weekes, T.C. et al. (Whipple Collab.): 1989, ApJ 342, 379
  \bibitem{} Weekes, T.C.: 2008, AIPC 1085, 3
\end{thebibliography}
\end{document}